\documentclass[aps,prd,twocolumn,superscriptaddress,showpacs,preprintnumbers,nofootinbib,notitlepage,longbibliography]{revtex4-2}
\usepackage{graphicx}
\usepackage{bm}
\usepackage[utf8]{inputenc}
\usepackage[hyperfootnotes=false]{hyperref}
\usepackage{multirow}
\usepackage{amsmath}
\usepackage{amsfonts}
\usepackage{amssymb} 
\usepackage{color}
\usepackage{xcolor}
\usepackage{xspace}
\usepackage{braket}
\usepackage{units}
\usepackage{slashed}
\usepackage{multirow}
\usepackage[normalem]{ulem}
\usepackage{notes2bib}

\graphicspath{ {plots/} }

\newcommand{\be}{\begin{equation}}
\newcommand{\ee}{\end{equation}}

\usepackage{bm} 

\renewcommand{\Re}{\ensuremath{\textrm{Re}}}

\newcommand{\mevnospace}{\ensuremath{{\mathrm{\,Me\kern -0.1em V}}}}
\newcommand{\gevnospace}{\ensuremath{{\mathrm{\,Ge\kern -0.1em V}}}}
\newcommand{\tevnospace}{\ensuremath{{\mathrm{\,Te\kern -0.1em V}}}}

\usepackage{mfirstuc} 
\newcommand{\addReviewer}[2]{
  \expandafter\newcommand\csname #1\endcsname[1]{{\bf \color{#2} \capitalisewords{#1}:\,##1}}
  \expandafter\newcommand\csname #1cor\endcsname[2]{{\color{#2} \capitalisewords{#1}:\,\st{##1}{\,\bf ##2}}}
  \expandafter\newcommand\csname #1color\endcsname{\,#2}
}

\usepackage{tikz, marginnote} 
\newcommand{\checkedby}[1]{
\ifdefined\CROSSCHECKS
  \marginnote{
    \begin{tikzpicture}
      \foreach \x [count=\xi] in {#1} {
         \node[shape=circle,inner sep=0mm,
         minimum size=2mm,
         fill=\csname \x color\endcsname] at (\xi*3mm,0) {};
       }
    \end{tikzpicture}
  }
\else
\fi
}

\usepackage{soul,color}
\definecolor{chromeyellow}{rgb}{1.0, 0.65, 0.0}
\definecolor{DodgeBlue}{rgb}{0.118, 0.565,1.000}
\definecolor{asparagus}{rgb}{0.53, 0.66, 0.42}
\definecolor{cadmiumgreen}{rgb}{0.0, 0.42, 0.24}

\addReviewer{AR}{asparagus}
\addReviewer{JR}{red}
\addReviewer{JRE}{blue}

\newcommand{\ucm}{Departamento de F\'isica Te\'orica and IPARCOS, 
Universidad Complutense de Madrid, 
E-28040 Madrid, Spain}

\begin{document}
\title{The unintuitive SU(3) flavor and chiral limits of hadron resonances}
\author{J.~R.~Pel\'aez}
\email{jrpelaez@ucm.es}\affiliation{\ucm}
\author{P. Rab\'an}
\email{praban@ucm.es}\affiliation{\ucm}
\author{J.~Ruiz de Elvira}
\email{jacobore@ucm.es}\affiliation{\ucm}
\preprint{IPARCOS-UCM-26-032}
\begin{abstract}
Contrary to naive expectations, poles used to define hadron resonances rigorously in the physical world may not evolve continuously to become degenerate in the SU(3)$_F$ and chiral limits of QCD. Instead, other shadow poles, usually ignored, may be the ones that degenerate and characterize the resonances in these limits. This feature is general, and we illustrate it first with the simple and familiar light-vector mesons, followed by the much-discussed light-scalar case. Their shadow poles and their degeneracy are found using the QCD low-energy effective theory unitarized to one loop.
\end{abstract}

\maketitle

{\it Introduction}---Resonances are ubiquitous in physics, from the classical forced damped oscillator to quantum spectra. In subatomic physics, they correspond to quasibound states characterized by a mass $M$ and width $\Gamma$, the inverse of their lifetime. Rigorously, they are defined by poles of scattering amplitudes at complex energies, i.e., $\sqrt{s_p}=M-i\Gamma/2$, where $s_p$ is the pole position in the energy-squared Mandelstam variable $s$. For narrow and isolated states, such poles generate the familiar Breit–Wigner (BW) peak $\sim M\Gamma/(M^2-s-iM\Gamma)$.
However, peaks are process-dependent, and BW parameterizations are not generically valid. Thus, pole definitions are increasingly adopted in the Review of Particle Physics (RPP, \cite{ParticleDataGroup:2024cfk}) as the standard characterization of resonances.

Quark flavor symmetry is a cornerstone of hadron physics and Quantum Chromodynamics (QCD). Although explicitly broken by quark masses, SU(3)$_F$ remains a useful approximate symmetry for the light quarks $u$, $d$, and $s$, providing the ground for hadron classification and interaction models. Yet, the multiplet assignment of many resonances remains controversial and often assumed rather than demonstrated. This occurs for the lightest scalar mesons studied here. 

The chiral limit of vanishing $u,d,s$ quark masses is of additional interest, since light-hadron masses and spontaneous SU(3) chiral-symmetry breaking in QCD are primarily governed by nonperturbative dynamics, while quark masses only provide explicit symmetry breaking. The mass gap between the pion, kaon, and eta mesons and the rest of the hadrons is due to their role as Nambu–Goldstone bosons of the spontaneous breaking.

Here we address the so-called Oakes-Yang problem, predating even the quark model~\cite{Oakes:1963zz},  of whether resonances---defined by their poles---belong to the same multiplet in the flavor-symmetric limit.  Surprisingly, we show that, in certain relevant cases, the poles that define resonances in our physical world do not become degenerate upon symmetry restoration. We first illustrate this behavior for the familiar light vectors and then argue that it is a generic phenomenon when poles lie between two thresholds that merge below the resonance in the symmetric limit. Finally, we verify that the much-debated light-scalar octet does become degenerate in the SU(3)$_F$ limit, but not continuously from the poles associated with the physical world. Our results challenge the common wisdom that the full dynamical and symmetry-breaking information of a resonance is encoded in a single pole.

{\it Formalism}---When the problem was posed first by Oakes and Yang, nothing was known about quarks---nor QCD---and the role of their masses in the symmetry breaking. Now, we can reach the SU(3)$_F$ limit by starting from the meson-meson scattering partial waves obtained from SU(3)$_F$ chiral perturbation theory (ChPT)~\cite{Gasser:1984gg}. This is the low-energy effective theory of QCD, written as a model-independent and systematic power expansion of the momenta and masses of the $\pi, K$ and $\eta$ mesons alone. We work in the isospin limit. Light vector and scalar resonances are generated,  without a priori assumptions on their existence or nature, by unitarizing these amplitudes with the next-to-leading order (NLO) coupled-channel inverse amplitude method (IAM)~\cite{GomezNicola:2001as,Pelaez:2004xp}.
This approach, not aimed at precision~\cite{Pelaez:2021dak}, reproduces the NLO ChPT expansion and describes fairly well the data on partial waves $t_J^{(I)}(s)$ of definite isospin $I$ and angular momentum $J=0,1$ up to 1.2 GeV.
The IAM requires only the fitting of the ChPT NLO-renormalized low-energy constants, without spurious parameters such as cutoffs or subtraction constants. The QCD quark-mass dependence up to NLO appears through the $m_\pi,m_K$ and $m_\eta$ masses, not only through kinematics, but also in the dynamics via the ChPT Lagrangian. 
The resulting $m_\pi$ and $m_K$ dependence of the light resonances appearing in elastic meson-meson scattering~\cite{Hanhart:2008mx,Nebreda:2010wv,Molina:2020qpw,Niehus:2020gmf}, is in fair agreement with later lattice calculations~\cite{Guo:2018zss,Mai:2019pqr,Andersen:2018mau,ExtendedTwistedMass:2019omo,Fischer:2020yvw,Rodas:2023gma,Boyle:2024hvv}. We will study the SU(3)$_F$ limit by continuously decreasing the strange quark mass in the NLO ChPT meson-mass expansion~\cite{Gasser:1984gg}, until it reaches the non-strange average value. This is equivalent to decreasing $m_K$ and $m_\eta$  until $m_K=m_\eta=m_\pi^{\text{phys}}$. From there, we will study the chiral limit. The unintuitive pole behavior we find is rather general and was suggested as a possibility for narrow resonances in simple multichannel meson-baryon scattering models before quarks, and QCD were proposed~\cite{Eden:1963zz,Eden:1964zz,Dalitz:1963ek,Ross:1963lhp}.
We show that it indeed occurs in a realistic description of meson-meson QCD dynamics, both for familiar one-channel resonances and non-standard states in multichannel scenarios.

{\it  Light vector mesons}---The familiar $\rho(770)$, $K^*(892)$, $\omega(782)$, and $\phi(1020)$ resonances are undoubtedly accepted to form an SU(3)$_F$ octet and a singlet. The first two are observed as well-isolated BW-like resonances in the elastic scattering of $\pi\pi$ and $K\pi$, respectively, to which they decay at 100\%. The second two are a mixture of the singlet and octet states $\phi_1$ and $\phi_8$, although in NLO two-meson scattering the former is decoupled~\cite{Oller:1999ag}. 
For brevity, let us label the (antisymmetric) octet partial waves by the resonance that dominates each, namely:
\begin{equation}
   \hat t_\rho\equiv\sigma_{\pi\pi}\, t_{1,8_a}^{(1)}\;\;
    \hat t_{K^*}\equiv\sigma_{K\pi}\, t_{1,8_a}^{(1/2)},\;\;
  \hat t_{\phi_8}\equiv\sigma_{KK}\, t_{1,8_a}^{(0)}.
\end{equation}
See App.\ref{app:ts} for details. For easier comparison, we multiply by $\sigma_{PQ}(s)=2k_{PQ}/\sqrt{s}$, where $k_{PQ}$ is the CM momentum of the lightest meson-meson state they couple to. 

\begin{figure}
\centering
\includegraphics[width=0.48\textwidth]{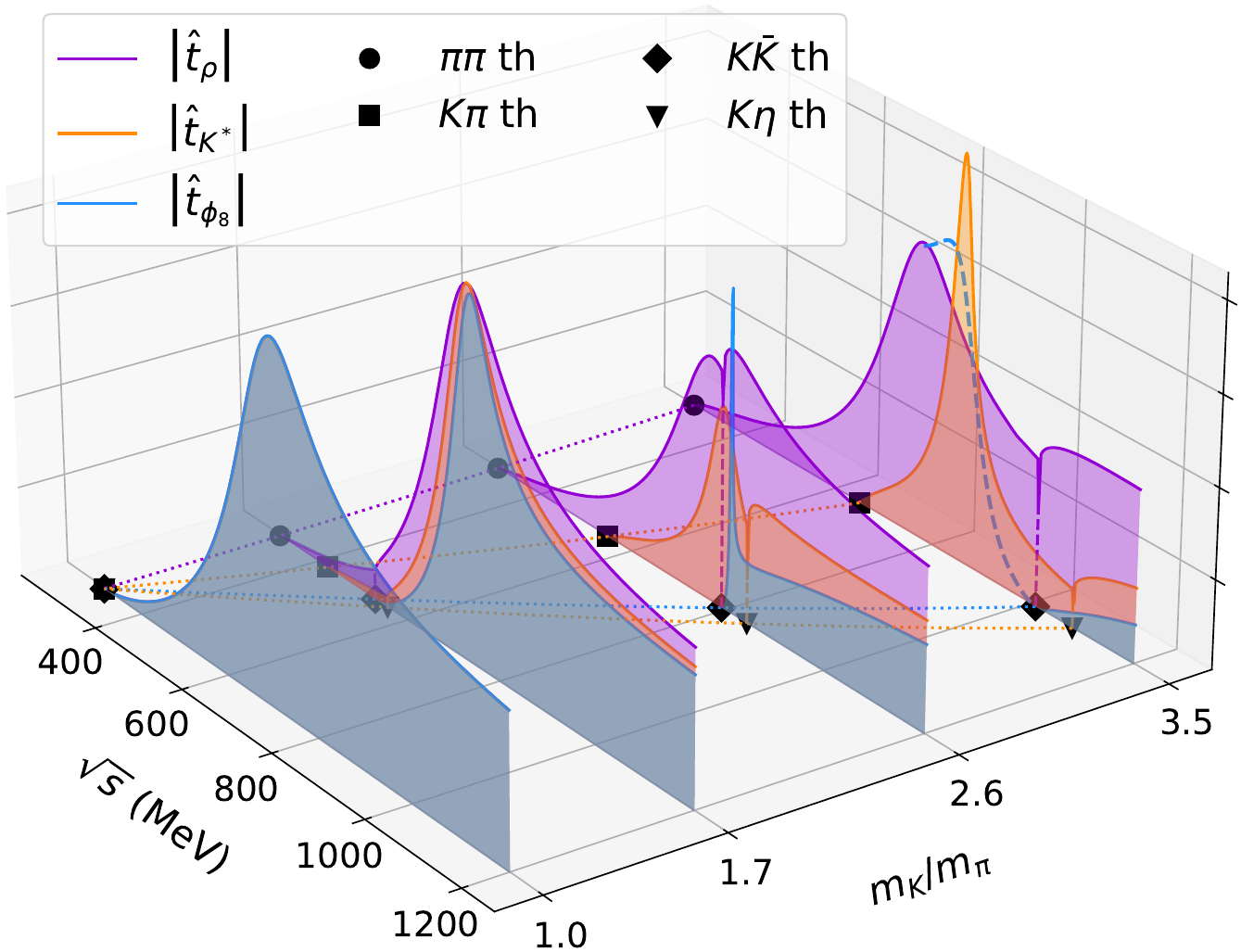}
\caption{ \small \label{fig:vectors_real_axis} 
ChPT-IAM vector partial waves on the $\sqrt{s}$ real axis, at representative $m_K/m_\pi$ values towards the SU(3)$_F$ limit.  The familiar $\rho$, $K^*$, and $\phi_8$ peaks get more alike until they merge when $m_K=m_\eta=m_\pi$. }
\end{figure}

In Fig.~\ref{fig:vectors_real_axis}, we show these three amplitudes above their lowest threshold, at four representative $m_K/m_\pi$ values, with $m_\pi\sim140\,$MeV fixed. For physical $m_K/m_\pi\simeq3.5$, we observe the familiar BW-like $\rho(770)$ and $K^*(892)$ peaks, whereas the $\phi_8$ appears below the $K\bar K$ threshold~\cite{Oller:1999ag}.  As $m_K/m_\pi$ decreases, the three peaks become more and more alike until they merge into a single peak when $m_K/m_\pi=1$. Similarly, the $K\pi$, $K\bar K$, and $K\eta$ thresholds (th) decrease until they coincide with that of $\pi\pi$. Moreover, not only the peaks, but the three amplitudes become one, as expected in the flavor limit.

\begin{figure*}
\centering
\includegraphics[width=\textwidth]{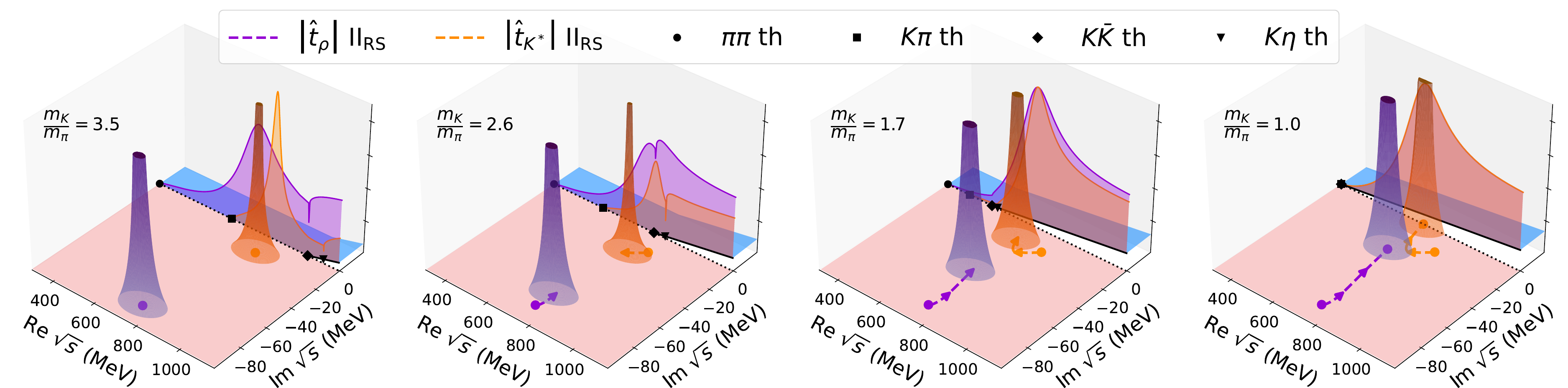}
\caption{ \small \label{fig:RSII} 
The leftmost panel shows the poles characterizing the physical $\rho(770)$ and $K^*(892)$, while the rest show
how their positions change for the same representative $m_K/m_\pi$ values of Fig.~\ref{fig:vectors_real_axis}. These poles are located in the second Riemann sheet (light red surface) of the lower-half complex $\sqrt{s}$-plane. The light-blue surface is the upper-half plane of the first sheet. 
Dashed lines with arrows follow the pole movements. To simplify the view, we only show 
$\vert \hat t_\rho\vert$ and $\vert \hat t_{K^*}\vert$ in a small domain around each pole and on the real axis (with different scales) as reached from the first sheet. Note the discontinuity gap between the first and second sheet growing from left to right until the second sheet is completely disconnected from the first when $m_K=m_\pi$.
}
\end{figure*}

However, resonances are not characterized by their process-dependent shape on the real axis, but by their process-independent associated poles. 
Let us first study the $\rho(770)$ and $K^*(892)$, and discuss the $\phi_8$ later. In Fig.~\ref{fig:RSII}, we show their poles in the complex $\sqrt{s}$ plane for the same representative values of $m_K/m_\pi$ used in Fig.~\ref{fig:vectors_real_axis}. For reference, we still plot on the real axis the modulus of the partial waves already shown in Fig.~\ref{fig:vectors_real_axis}. 
In the far-left panel, corresponding to the physical case $m_K/m_\pi\simeq3.5$, we find the familiar, narrow, and isolated $\rho(770)$ and $K^*(892)$ poles dominating, respectively, the $\vert \hat t_\rho\vert$ and $\vert \hat t_{K^*}\vert$ elastic regions.
In the panels to the right, the $\rho$ pole moves closer to the real axis, barely changing its pole mass, whereas the $K^*$ pole gets slightly lighter, with its pole width first increasing a little and then decreasing. As intuitively expected, the $\rho$ and $K^*$ poles seem to get closer as $m_K/m_\pi$ decreases, with $m_\pi$ fixed.

Surprisingly, \textit{against that intuition, the $\rho$ and $K^*$ poles do not degenerate into a single pole in this SU(3)$_F$-symmetric regime}, as shown in the far-right panel of Fig.~\ref{fig:RSII}, where $m_K/m_\pi=1$. There, we still see two poles. Why do they not degenerate into one? Moreover, they are both much closer to the real axis than in the physical case---the $K^*$ pole even sits right on the real axis. Although this should lead to a much narrower amplitude peak, it remains almost as wide as in the physical case.
What causes this ``two non-degenerate narrow-poles" versus ``only one wide-peak" mismatch? 

There are two reasons: first, the familiar poles characterizing physical light vectors lie in a Riemann sheet that becomes disconnected from the first through the real axis in the symmetric limit. Second, additional ``shadow" poles, customarily neglected for elastic resonances, are present in the sheet connected to the amplitude in the symmetric limit. Let us discuss these features in order.

\begin{figure}
\centering
\includegraphics[width=0.48\textwidth]{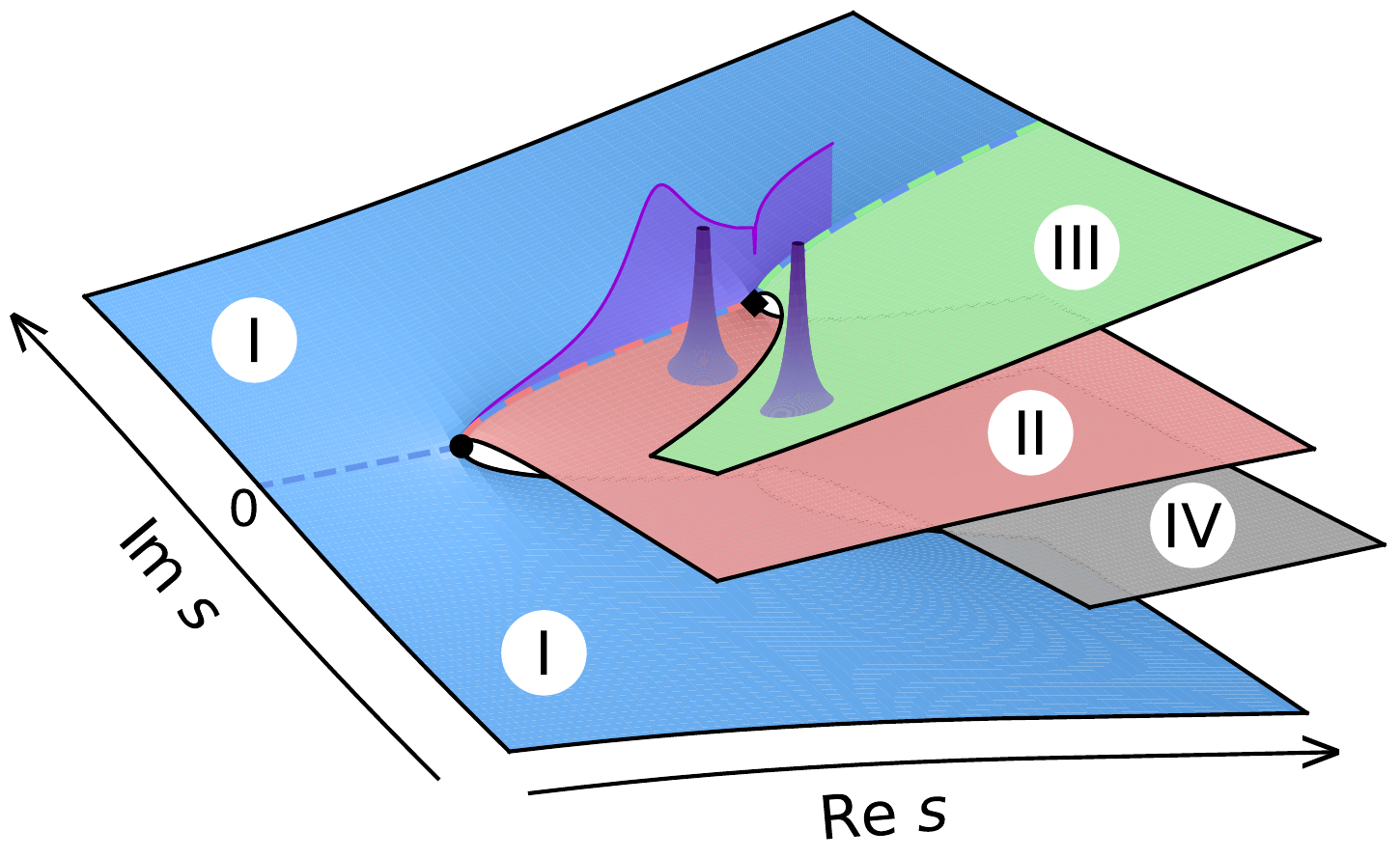}
\caption{\label{fig:sheets} Sheet structure for two thresholds. A branch cut extends along the real axis from each threshold to $+\infty$. The physical amplitude $\vert \hat{t}_\rho \vert$ lies on the upper-half plane of sheet I. Sheets I and II are adjacent only between the first and second thresholds; beyond the second threshold, sheet III becomes the adjacent sheet. The conventional resonance pole lies on sheet II, while a shadow pole resides on sheet III.}
\end{figure}

{\it  Multiple Riemann sheets}---In Fig.~\ref{fig:sheets}, we illustrate the sheet structure of scattering amplitudes in the complex $s$-plane (valid also for the $\Re \sqrt{s}\geq0$ half-plane), when two thresholds are present. The number of sheets doubles with each additional threshold. For $\hat t_{\rho}$, the first threshold is $\pi\pi$ and the second $K\bar K$. For $\hat t_{K^*}$, the first is $K\pi$ and the second $K\eta$.  The physical amplitude lies on the real axis as approached from the upper-half plane of the first or ``physical" sheet (blue), which is free of resonance poles.\bibnote{For other hadronic systems, poles associated with bound states may appear in the first sheet in the real axis below the first threshold. For meson-meson scattering, this happens, for instance, if $m_\pi$ is increased~\cite{Hanhart:2008mx}.} 
Each threshold marks the appearance of a cut discontinuity when crossing the real axis. Depending on whether this crossing is made continuously or not, one gains access to a different sheet. The generalization to $N$ thresholds and $2^N$ sheets is straightforward.

The second sheet (red) lower half-plane is reached by 
analytically continuing the amplitude in the first sheet through the first cut. It is thus continuously connected, or adjacent, to the upper half-plane of the first sheet, but only in the real segment between the first and second thresholds. 
When a pole sits on the second sheet near that segment, it produces the characteristic resonance peak in $\vert t\vert$ on the real axis. This setup is sketched in Fig.~\ref{fig:sheets} and corresponds to the physical BW-like $\rho(770)$ or $K^*(892)$ in the leftmost panel of Fig.~\ref{fig:RSII}. However, as $m_K$ decreases, the second threshold approaches the first, and the second sheet is connected to the first in a smaller segment. This is represented in Fig.~\ref{fig:RSII} as a white gap between the first and second sheets. This gap opens nearer and nearer the $\pi\pi$ threshold when moving from the left to the right panel, until, in the rightmost panel, the second sheet is completely disconnected from the first.  We thus say the familiar poles lie in the ``physically contiguous" but not in the ``symmetrically contiguous" sheet. 
Hence, in the $m_K,m_\eta \to m_{\pi}^{\text{phys}}$ limit, light-vector poles on the second sheet no longer determine the amplitude shape, no matter how close they are to the real axis. 
But then, how can resonant amplitudes become degenerate? This is an instance of the 1963 Oakes-Yang problem~\cite{Oakes:1963zz}.

{\it  Shadow poles}--- Soon after, the existence of resonance poles lying in additional sheets was proposed as a possible solution within simple models of multichannel systems~\cite{Dalitz:1963ek,Eden:1963zz,Eden:1964zz,Ross:1963lhp}. As shown in Fig.~\ref{fig:sheets}, above the second threshold, the adjacent sheet is the third (green), reached by crossing both the first and second cuts continuously. Poles in sheet III shape hadronic amplitudes below the second threshold only if they lie very close to it~\cite{Au:1986vs,Hanhart:2007yq,Zhang:2009bv,Kang:2016jxw,Wang:2024ytk,Asokan:2022usm,Zhang:2024qkg}. Otherwise, they are ignored, as for the $\rho(770)$ and $K^*(892)$, which are normally considered single-channel or ``elastic" resonances. 
 However, here we show that such ``shadow" poles exist and prove their degeneracy in the $m_K\to m_\pi^{\text{phys}}$ symmetric limit.
 
\begin{figure*}
\centering
\includegraphics[width=\textwidth]{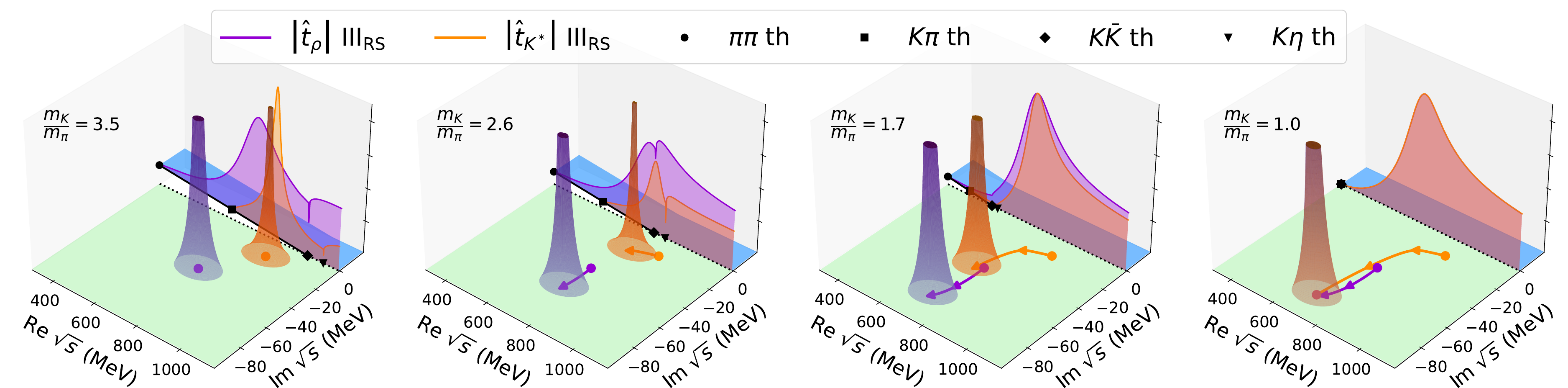}
\caption{ \small \label{fig:RSIII} 
Poles in the third Riemann sheet (green) associated with the $\rho(770)$ and $K^*(892)$, for the same representative $m_K/m_\pi$ values of Figs.~\ref{fig:vectors_real_axis} and~\ref{fig:RSII}. Solid lines with arrows follow the pole movements. The two poles degenerate in the light-flavor limit. To simplify the view, we only show 
$\vert \hat t_\rho\vert$ and $\vert \hat t_{K^*}\vert$ in a small domain around each pole and on the real axis (with different scales). The light-blue surface corresponds to the upper-half plane of the first sheet. Note that, from left to right, the white gap between the first and third sheets keeps closing until, in the symmetric limit, they are contiguous from the only threshold.
}
\end{figure*}

Actually, we have found the light-vector third-sheet poles, shown for representative $m_K/m_\pi$ values in Fig.~\ref{fig:RSIII}. In the physical case (leftmost panel), they lie well below their corresponding second thresholds, where the third sheet is disconnected from the first, as also sketched in Fig.~\ref{fig:sheets} for the physical $\rho(770)$.
These shadow poles barely affect the amplitude, and it makes perfect sense to ignore them in real-world phenomenology.

However, because pseudo-Nambu-Goldstone boson masses depend more strongly on light-quark masses than resonances do, the second thresholds decrease faster than the resonance poles shift as $m_K\to m_\pi$ (see Fig.~\ref{fig:RSIII}). 
We have chosen the $m_K/m_\pi=2.6$ value, since it is when 
the $K\bar K$ threshold falls on top of the $\rho$ peak in the amplitude. Then, as seen in Figs.~\ref{fig:vectors_real_axis}, ~\ref{fig:RSII} and~\ref{fig:RSIII}, the presence of the threshold splits the peak in two, and cannot be described with a simple BW shape. The part of the peak below the threshold is due to the second-sheet pole, and the part above is due to the third-sheet shadow pole. 

As already discussed, as the second threshold decreases, sheet II becomes disconnected at lower energies, and its poles do not shape the amplitude in the real axis, whereas those in sheet III do. The shadow-pole role is exchanged. In the light-flavor limit, it is the third-sheet $\rho(770)$ and $K^*(892)$ poles that coalesce into a single position with the same residue, giving rise to a single amplitude peak.
This is how resonance poles from the same multiplet degenerate in the symmetric limit.

In hindsight, in the $m_K\to m_\pi^{\text{phys}}$ limit, since pairs made of pions, kaons, and etas are degenerate, it makes no sense to treat their cuts differently. If some of them are crossed continuously and others discontinuously, we break SU(3)$_F$ symmetry in the analytic continuation, even if the amplitudes are SU(3)$_F$ symmetric in the real axis. In that limit, the only relevant poles for the amplitude on the real axis are those in the sheet reached by crossing all cuts continuously, i.e., the sheet we have called ``symmetrically-contiguous". For real-world vector mesons, these are shadow poles.

\begin{figure}
\centering
\includegraphics[width=0.48\textwidth,height=4cm]{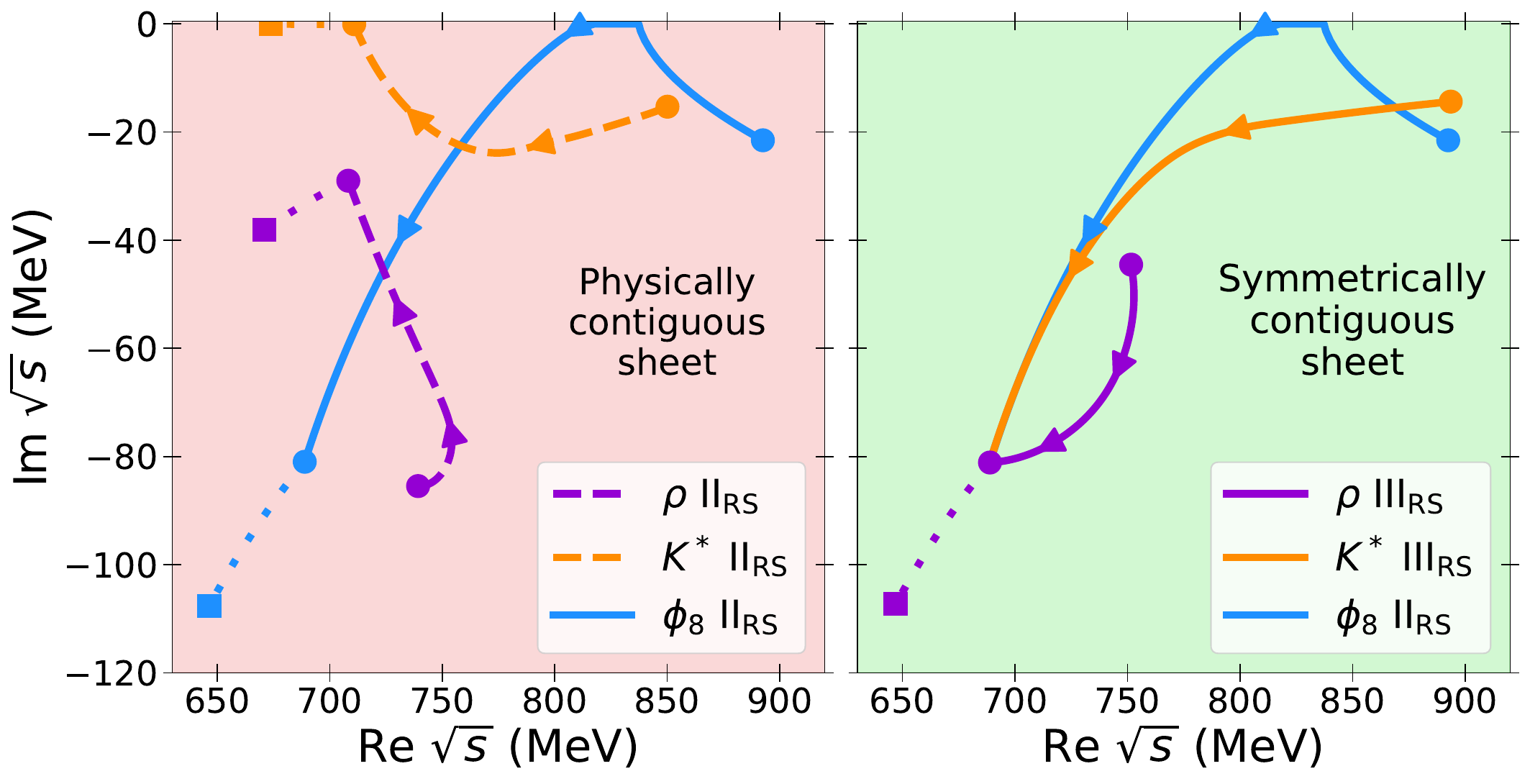}
\caption{ \small \label{fig:vectors_trajectories} 
Light-vector pole trajectories towards the SU(3)$_F$ limit $m_K\to m_{\pi}^{\text{phys}}$, with $m_\pi$ fixed, both in their physically-(left)
and symmetrically-contiguous sheets (right), which are the same for the $\phi_8$.  Only poles in their symmetrically contiguous sheet degenerate in the symmetric limit. From there, dotted lines represent their paths to the chiral limit. }
\end{figure}

{\it The SU(3)$_F$ and chiral limits of light-vector poles}--- Fig.~\ref{fig:vectors_trajectories} shows the $\rho$ and $K^*$ pole movements in detail in the second and third sheets. We have now added the $\phi_8$ pole trajectory. In two-meson scattering below 1.2 GeV, its partial wave couples only to $K\bar K$ and has one cut and two sheets, so that the second sheet is both physically and symmetrically contiguous. Note that the $\phi_8$ pole moves briefly on the real axis, where it meets the left cut, extending from $4(m^2_K-m_\pi^2)$ to $-\infty$. 
 
In addition, in Fig.~\ref{fig:vectors_trajectories}, pole trajectories are extended as dotted lines to reach the chiral limit $m_\pi\to0$, keeping the light-flavor symmetry. With vanishing quark masses, the resonance mass arises solely from non-perturbative QCD dynamics. Since the common multiplet mass gets reduced by about 50 MeV, but the two-NGB threshold is $\sim 280\,$MeV lower, the available phase space is larger, and the resonance width increases (see~\cite{Hanhart:2008mx}).

Hence, by continuously restoring light-flavor symmetry, we have shown that the poles associated with the light-vector octet indeed degenerate into a single pole. However, as $m_K\to m_\pi^{\text{phys}}$, this only occurs for their shadow poles, not for the poles used to characterize them in the real world. Of course, no one questions this octet, but it helps us to understand other contentious cases. 

{\it The controversial light-scalar mesons}---The assignment of the $f_0(500)/\sigma$, $K^*(700)/\kappa$, $f_0(980)$, and $a_0(980)$ multiplet remains disputed (see ``Scalar Mesons below 1 GeV" in~\cite{ParticleDataGroup:2024cfk}). The last two are narrow and twice as heavy as the first one, requiring a coupled-channel formalism due to their proximity to the $K\bar K$ threshold. In contrast, the two lightest have widths comparable to their masses, couple mostly to a single channel, and do not appear as amplitude peaks. Thus, they were not identified as resonances for decades.
As simple BW parameterizations fail, they were 
only recently accepted in the RPP following rigorous data-driven dispersive determinations

In this work, we demonstrate how these states degenerate into one octet and one singlet, in the $m_K\to m_\pi^{\text{phys}}$ flavor limit. Once again, this degeneracy is not reached by their ``physical poles," but by their shadow poles. As shown below, the Riemann sheet structure is now more intricate, pole trajectories are more involved, and the mixing of the octet $f_8$ and singlet $f_1$ persists until the light-flavor limit is attained.

Once again, we label each (symmetric) octet scalar partial wave by the resonance that dominates it:
\begin{equation}
\hat t_{a_0}\equiv \sigma_{\pi\eta}\,  t_{0,8_s}^{(1)},\;
  \hat t_{\kappa}\equiv\sigma_{K\pi}\,   t_{0,8_s}^{(1/2)},\;
 \hat t_{f_8}\equiv \sigma_{\pi\pi}\, t_{0,8_s}^{(0)},
  \end{equation}
and similarly for the singlet combination $\hat t_{f_1}\equiv\sigma_{\pi\pi}\,t_{0,1}^{(0)}.$

\begin{figure}
\centering
\includegraphics[width=0.48\textwidth]{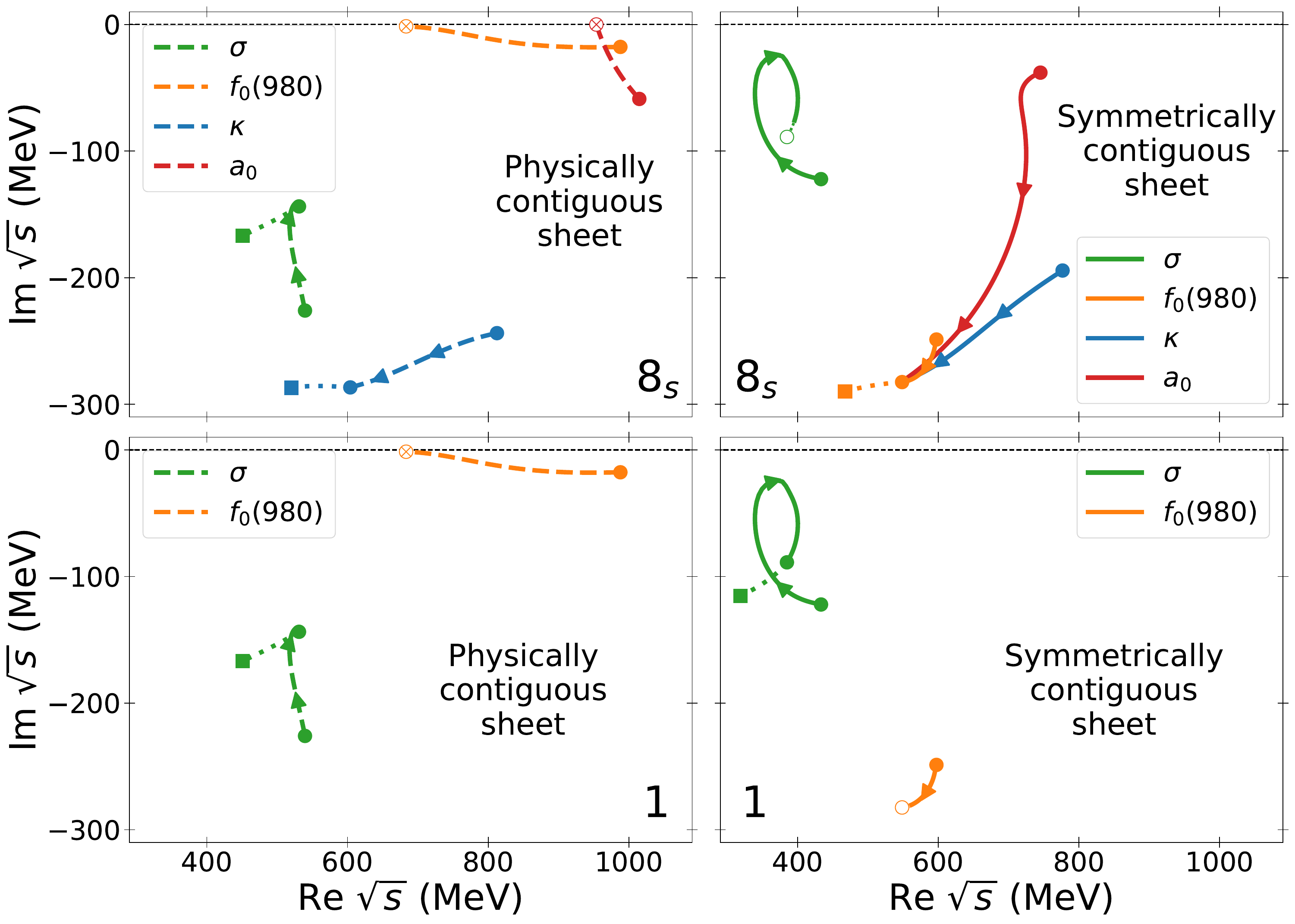}
\caption{\small SU(3)$_F$-restoring pole trajectories in the physically- (left) and symmetrically-contiguous (right) sheets, for the octet (up) and singlet (down) scalar amplitudes. The octet poles degenerate in the symmetric limit only in the symmetrically contiguous sheet. The dotted lines represent the flavor-symmetric trajectories towards the chiral limit.
\label{fig:scalarpolesmove} 
}
\end{figure}

Note that both the singlet and the octet now couple to three different meson-meson states $\pi\pi$, $K\bar K$, and $\eta\eta$. This means three different thresholds, three cuts, and eight Riemann sheets, which we do not attempt to sketch. As before, we use NLO ChPT amplitudes unitarized with the IAM to determine poles.
We are only interested in poles lying on the physically and symmetrically contiguous sheets, whose motion is shown in Fig.~\ref{fig:scalarpolesmove} 

For light-scalar poles, the physically connected sheet is the second~\bibnote{Even the $f_0(980)$ and $a_0(980)$ characteristic shapes near the $K\bar K$ threshold are due to poles in this sheet (see~\cite{Wang:2022vga,Burkert:2022bqo})}, which we show in the left panels of Fig.~\ref{fig:scalarpolesmove}. This time, not only do these poles fail to degenerate in the light-flavor or chiral limits, but some even disappear. As expected, the $a_0(980)$ and $K_0^*(700)/\kappa$ arise only in the octet amplitude.
Surprisingly, in this sheet, both the $f_0(500)/\sigma$ and $f_0(980)$ poles appear simultaneously in the octet and singlet amplitudes, even in the flavor-symmetric limit.
The octet and singlet can still mix because, to reach this sheet, we break flavor symmetry by crossing the first cut continuously, but not the other two.  

The flavor and chiral limits are only intuitive for shadow poles in the symmetrically contiguous sheet (right column of Fig.~\ref{fig:scalarpolesmove}). This is again the third sheet for the $a_0(980)$ and $K^*_0(700)/\kappa$, but it is the fifth sheet for the $f_0(500)/\sigma$ and $f_0(980)$, which appear in a three-threshold case. 
Due to mixing, $\sigma$ and $f_0(980)$ poles show up in both the octet and singlet amplitudes, but only until the symmetry is restored.
Then, the one evolving from the $f_0(980)$ decouples from the singlet channel, whereas the one coming from the $f_0(500)/\sigma$ (green) decouples from the octet. Remarkably, all poles except the latter degenerate into a common octet pole (upper right panel). We have thus shown that the light-meson poles are indeed a continuous deformation of a properly defined octet and a singlet in the flavor-symmetric limit. Note that both remain massive in the chiral limit. 

{\it Discussion}--- Previous degeneracy studies approach the symmetric limit by increasing $m_\pi$~\cite{Wali:1964xv,Oller:2003vf,Roca:2005nm,Jido:2003cb,Bruns:2021krp,Yamaguchi:2016kxa} (or the $u,d$ quark masses), sometimes decreasing $m_K$ and $m_\eta$ too. While useful for lattice QCD comparison, this strategy precludes studying the chiral limit and worsens chiral convergence~\cite{Yamaguchi:2016kxa}. When $m_\pi$ is made heavy enough to lift the lowest threshold above the resonance masses, ``physical'' poles may degenerate directly as bound states on the first Riemann sheet, bypassing shadow poles. Although not interested in such a limit, we have verified that this occurs for light vectors, whereas some shadow poles remain essential for certain multiplets in meson-baryon scattering~\cite{Wali:1964xv,Bruns:2021krp}.

By increasing $m_\pi$ in unitarized LO chiral amplitudes, prior works investigated multiplet degeneracy for light-scalar~\cite{Oller:2003vf}, pseudovector~\cite{Roca:2005nm}, and baryon resonances~\cite{Jido:2003cb}. While their resulting trajectories appear smooth, many hide a subtle artifact: they initially track the physical pole but at an intermediate point jump to a shadow pole on the symmetrically contiguous sheet. This intuitive but discontinuous patching contrasts with our approach, which is rigorously continuous, but may be unintuitive.

{\it Summary}---We have demonstrated that, contrary to intuition, the poles that rigorously define hadron resonances within the same $\text{SU}(3)_F$ multiplet do not necessarily degenerate as the flavor-symmetric or chiral limits are continuously approached. When this happens, the degeneracy is realized by shadow poles, even if they are negligible in real-world phenomenology. We have shown this behavior for the familiar light-vector octet and the controversial light-scalar mesons. Still, this mechanism is generic whenever a physical resonance lies between two thresholds that, in the symmetric limit, become degenerate below its pole mass. 
Consequently, the information about the symmetry-breaking dynamics of a resonance is not encoded solely in the pole used to define it physically.
This observation must be taken into account for the identification of multiplets of any symmetry and, particularly, for a rigorous spectroscopic classification of hadrons.

{\it Acknowledgments}---
This work is part of the Grant PID2022-136510NB-C31, funded by MCIN/AEI/10.13039/501100011033. It has also received funding from the European Union’s Horizon 2020 research and innovation program under Grant Agreement No.824093. P. R. is supported
by the MIU (Ministerio de Universidades, Spain) fellowship FPU21/03878, and J.R.E. by the Ram\'on y Cajal program (RYC2019-027605-I) of the Spanish MICIU.

\appendix

\section{Octet and singlet partial wave combinations}
\label{app:ts}

Although they are straightforward to calculate, for completeness, we provide the SU(3)$_F$ octet and singlet combinations, in the isospin limit, of two-body scattering partial waves of pions, kaons, and etas, relevant for the main text. They are just combinations of partial waves $t^{(I)}_J(s)$, of definite isospin $I$ and angular momentum $J$, obtained using the Clebsch-Gordan coefficients for SU(3).

The partial-wave amplitude is related to the S matrix $S^{(I)}_{J,a\to b}$, of definite isospin and angular momentum, of the $a\to b $ process through
\begin{equation}
	S^{(I)}_{J,a\to b}(s)=\delta_{ab}+2i\sqrt{\sigma_a(s)\sigma_b(s)}\,t^{(I)}_{J,a\to b}(s).
	\label{eq:unit_pw_S}
\end{equation}

For the vector case, $J=1$, Bose symmetry requires an antisymmetric flavor state, which can only be an octet. The partial waves read:
\begin{align}
	t_{1,8_a}^{(1)}=&\tfrac{4}{3}t_{1,\pi\pi}^{(1)}+\tfrac{2}{3}t_{1,K \bar K}^{(1)}+\tfrac{4\sqrt{2}}{3}t_{1,\pi\pi\to K\bar K}^{(1)}\equiv \hat t_\rho/\sigma_{\pi\pi},\label{eq:trho}\\
	t_{1,8_a}^{(1/2)}=&t_{1,K \pi}^{(1/2)}+t_{1,K \eta}^{(1/2)}+2t_{1,K\pi\to K\eta}^{(1/2)}\equiv \hat t_{K^*}/\sigma_{K\pi},\label{eq:tKstar}\\
	t_{1,8_a}^{(0)}=&2t_{1,K \bar K}^{(0)}\equiv \hat t_{\phi_8}/\sigma_{KK}.
	\label{eq:tphi8}
\end{align}

For the scalars, $J=0$, there is a symmetric octet, which reads
\begin{align}
	t_{0,8_s}^{(1)}=&\tfrac{4}{5}t_{0,\pi\eta}^{(1)}+\tfrac{6}{5}t_{0,K \bar K}^{(1)}-\tfrac{4\sqrt{6}}{5}t_{0,\pi\eta\to K\bar K}^{(1)}\equiv \hat t_{a_0}/\sigma_{\pi\eta},\label{eq:ta0}\\
	t_{0,8_s}^{(1/2)}=&\tfrac{9}{5}t_{0,K \pi}^{(1/2)}+\tfrac{1}{5}t_{0,K \eta}^{(1/2)}-\tfrac{6}{5}t_{0,K\pi\to K\eta}^{(1/2)}\equiv \hat t_{\kappa}/\sigma_{K\pi},\label{eq:kappa}\\
	t_{0,8_s}^{(0)}=&\tfrac{6}{5}t_{0,\pi\pi}^{(0)}+\tfrac{2}{5}t_{0,K \bar K}^{(0)}+\tfrac{2}{5}t_{0,\eta\eta}^{(0)}-\tfrac{4\sqrt{3}}{5}t_{0,\pi\pi\to K\bar K}^{(0)}\nonumber\\
	&+\tfrac{4\sqrt{3}}{5}t_{0,\pi\pi\to \eta\eta}^{(0)}-\tfrac{4}{5}t_{0,K\bar K\to\eta\eta}^{(0)}\equiv \hat t_{f_8}/\sigma_{\pi\pi},\label{eq:tf8}
\end{align}
as well as a singlet combination:
\begin{align}
	t_{0,1}^{(0)}=&\tfrac{3}{4}t_{0,\pi\pi}^{(0)}+t_{0,K \bar K}^{(0)}+\tfrac{1}{4}t_{0,\eta\eta}^{(0)}+\sqrt{3}t_{0,\pi\pi\to K\bar K}^{(0)}\nonumber \\
	&-\tfrac{\sqrt{3}}{2}t_{0,\pi\pi\to \eta\eta}^{(0)}-t_{0,K\bar K\to\eta\eta}^{(0)}\equiv \hat t_{f_1}/\sigma_{\pi\pi}.\label{eq:tf1}
\end{align}
For easier comparison in the figures in the main text, we have defined $\hat t$ partial waves by dividing each partial-wave combination above by the phase space of the lightest two-meson channel it couples to.


\bibliographystyle{apsrev4-2}
\bibliography{largebiblio.bib}

\end{document}